\documentclass[english]{article}	

\usepackage[utf8]{inputenc}				
\usepackage[big,online]{dgruyter}	
\usepackage{lmodern} 
\usepackage{microtype}
\usepackage[modulo]{lineno}
\modulolinenumbers[1]
\usepackage[backend=biber,
    style=numeric,
    natbib=true,
    url=false, 
    doi=true,
    sorting=none,
    eprint=false]{biblatex}
\usepackage{siunitx}
\DeclareSIUnit\Gbps{Gbps}
\usepackage{hyperref}
\usepackage{graphicx}
\graphicspath{{figures/}}
\usepackage{authblk}
\usepackage{amsmath}

\begin{document}

\title{Artificial optoelectronic spiking neuron based on a resonant tunnelling diode coupled to a vertical cavity surface emitting laser}

\author[1]{Mat\v{e}j Hejda}
\author[2]{Ekaterina Malysheva}
\author[1]{Dafydd Owen-Newns} 
\author[3]{Qusay Raghib Ali Al-Taai}
\author[1]{Weikang Zhang} 
\author[4]{Ignacio Ortega-Piwonka}
\author[4]{Julien Javaloyes}
\author[3]{Edward Wasige}
\author[2]{Victor Dolores-Calzadilla}
\author[5]{José M. L. Figueiredo}
\author[6]{Bruno Romeira}
\author[1]{Antonio Hurtado}
\runningauthor{M.~Hejda et al.}

\affil[1]{\protect\raggedright 
Institute of Photonics, SUPA Dept of Physics, University of Strathclyde, Glasgow, United Kingdom, e-mail: matej.hejda@strath.ac.uk, dafydd.owen-newns@strath.ac.uk, weikang.zhang@strath.ac.uk, antonio.hurtado@strath.ac.uk}
\affil[2]{\protect\raggedright 
Eindhoven Hendrik Casimir Institute, Eindhoven University of Technology, Eindhoven, The Netherlands, e-mail: e.malysheva@tue.nl, v.calzadilla@tue.nl}
\affil[3]{\protect\raggedright 
High Frequency Electronics Group, University of Glasgow, Glasgow, United Kingdom, e-mail: qusayraghibali.al-taai@glasgow.ac.uk, edward.wasige@glasgow.ac.uk}
\affil[4]{\protect\raggedright 
Dept de Física and IAC-3, Universitat de les Illes Balears, Palma de Mallorca, Spain, e-mail: ignacio.ortega.piwonka@gmail.com, Julien.Javaloyes@uib.es}
\affil[5]{\protect\raggedright 
Centra-Ci\^{e}ncias and Departamento de F\'{i}sica, Faculdade de Ci\^{e}ncias, Universidade de Lisboa, Lisboa, Portugal, e-mail: jmfigueiredo@fc.ul.pt}
\affil[6]{\protect\raggedright 
INL – International Iberian Nanotechnology Laboratory, Ultrafast Bio- and Nanophotonics Group, Braga, Portugal, e-mail: bruno.romeira@inl.int}
	
\abstract{Excitable optoelectronic devices represent one of the key building blocks for implementation of artificial spiking neurons in neuromorphic (brain-inspired) photonic systems. This work introduces and experimentally investigates an opto-electro-optical (O/E/O) artificial neuron built with a resonant tunnelling diode (RTD) coupled to a photodetector as a receiver and a vertical cavity surface emitting laser as a the transmitter. We demonstrate a well defined excitability threshold, above which this neuron produces $\sim$\SI{100}{\nano\second} optical spiking responses with characteristic neural-like refractory period. We utilise its fan-in capability to perform in-device coincidence detection (logical AND) and exclusive logical OR (XOR) tasks. These results provide first experimental validation of deterministic triggering and tasks in an RTD-based spiking optoelectronic neuron with both input and output optical (I/O) terminals. Furthermore, we also investigate in theory the prospects of the proposed system for its nanophotonic implementation with a monolithic design combining a nanoscale RTD element and a nanolaser; therefore demonstrating the potential of integrated RTD-based excitable nodes for low footprint, high-speed optoelectronic spiking neurons in future neuromorphic photonic hardware.}


\maketitle

\section{Introduction}
Artificial Intelligence (AI) and Machine Learning (ML) algorithms nowadays power a wide range of advanced computational tasks, ranging from natural language processing and realistic image synthesis \cite{Ramesh2021} to solutions solving major challenges such as protein folding \cite{Jumper2021}. As a general principle, it can be stated that higher computational capability of AI models goes hand in hand with their scale. Hence, further growth in the size of these models is expected as new, more complex tasks are being explored. This can be observed in current models such as GPT-3, which operates with 175 billion parameters \cite{Brown2020}, over a magnitude increase in the number of parameters when compared to its previous iteration. With the growth of scale and increase in resource and requirements of these AI models, the chips on which those algorithms run come more into the spotlight, fuelling the search for alternative, AI-optimised hardware. In particular, approaches beyond the conventional Von-Neumann architecture of digital processors (with distinct memory and logic units) are receiving increasing interest. These alternative computing schemes offer the promise of relieving the stalling chip performance improvements due to CMOS downscaling bottlenecks and architecture limitations. Neuromorphic (brain-inspired) engineering is a prime example of such unconventional computing approach. 

Neuromorphic computing systems attempt to harness the vast computational capabilities and power efficiency of the brain by mimicking and abstracting their architecture. These systems rely on very high degree of parallelism and concepts such as event-based asynchronous computation and in-memory computing. Driven both by their utility in the fields of AI and computational neuroscience, neuromorphic computers are being developed by both academic \cite{Neckar2019,Hoppner2021} and industrial parties \cite{DeBole2019,Orchard2021} in a variety of technology platforms. In particular, neuromorphic realisations based on photonics offer some highly desirable benefits. The use of light allows for high bandwidth and low-loss, wavelength-division multiplexed (WDM) communication schemes without unwanted inductive crosstalk and resistive heating in wires, while advances in the field of photonic integrated circuits (PICs) allow for high-density chip integration. The field is currently ongoing rapid expansion, with many different classes of photonic devices being investigated for brain-inspired computing and AI acceleration. These include quantum-dot lasers \cite{Robertson2018,Sarantoglou2020}, superconducting nanowires \cite{Schneider2018, Toomey2019a}, integrated photonic components including modulators \cite{George2019, Tait2019a, Mourgias-Alexandris2020a}, semiconductor optical amplifiers, \cite{Mourgias-Alexandris2019,Shi2021_APLPhotonics}, micro-rings \cite{Coomans2011,Xiang2020_MRR}, phase change material-based PICs \cite{Feldmann2019,Carrillo2019} as well as vertical cavity surface emitting lasers (VCSELs) subject to injection locking \cite{Hejda2020,Robertson2020} or with saturable absorber sections \cite{Xiang2020a,Zhang2019c}. Furthermore, VCSELs have been previously demonstrated as a viable technology for spike-based optical computing by utilising their high speed spiking dynamics for tasks such as all-optical convolution \cite{Zhang2021a}, pattern classification \cite{Robertson2020a} or rate-coded encoding of image data \cite{Hejda2021}. Simultaneously, a recent study has demonstrated VCSELs suitability for integration into arrays \cite{Heuser2020}, which is crucial for realisation of larger-scale on-chip integrated circuits.

In this work, we introduce a modular, excitable, optoelectronic spiking neuron based on a resonant tunnelling diode (RTD) electrically coupled to a PD (Thorlabs PDA8GS) serving as a receiver and to a telecom-wavelength, pigtailed VCSEL serving as a transmitter, together realising an O/E/O system. RTDs are a class of semiconductor devices that typically employ a double barrier quantum well (DBQW) semiconductor heterostructure, allowing for ultrafast quantum tunnelling through the resonant states of the well. This sets RTD-based oscillators among the fastest semiconductor devices operating at room temperature, with currently highest achieved frequency reaching \SI{2}{\tera\hertz} \cite{Izumi2017}. RTDs exhibit a highly nonlinear, \textit{N}-shaped I-V with a negative differential conductance (NDC) region, which introduces gain and nonlinear dynamical responses. Previously, RTDs have been successfully employed for photodetection with very high sensitivity \cite{Weng2015,Pfenning2016}, as reciever systems \cite{Nishida2019}, for small-scale THz imaging \cite{Miyamoto2016} and in circuits for high-speed data transmission \cite{Wang2018_84GHz,Zhang2019_79GHz}. RTD-based oscillator circuits can also exhibit excitability, yielding these devices as highly-promising elements for use in novel brain-inspired computing paradigms \cite{Romeira2013} and in so-called cellular neural networks \cite{Hanggi2001,Mazumder2009}. Previous works have demonstrated spike-shaped oscillations \cite{Al-Taai2021_EuMIC} and neuron-like, excitable stochastic (noise-driven) spiking in systems of RTDs connected to either a laser diode (RTD-LD) \cite{Romeira2013} or a photodetector (RTD-PD) and using electrical noise or modulated optical input \cite{Zhang2021_PDRTD}. A signal-regenerating spiking memory cell with an optoelectronic RTD circuit has also been reported \cite{Romeira2016} as well as optically induced stochastic resonance effects \cite{Hartmann2011}. Recently, nanoscale RTD-based LEDs \cite{Romeira2020} have been proposed as a viable solution for low-power, high-speed and low footprint optical spiking nodes. Numerically, the dynamics and spiking signal propagation characteristics were recently studied in a master-receiver (RTD-LD to PD-RTD) system \cite{Ortega-Piwonka2022_OME} and feed-forward spiking neural network based on RTD-LD/PD master-receiver nodes, with spatiotemporal spike pattern detection and information processing functionality \cite{Hejda2022_PRAppl}. 

In Section \ref{sec:layout}, we introduce the layout of the studied RTD-based optoelectronic node and the experimental setup. While previous studies focused on stochastic, noise-induced spiking \cite{Zhang2021_PDRTD}, our layout enables deterministic, user controlled spike triggering of the RTD-based photonic neuron, which is key for practical information processing. Furthermore, our photonic neuron combines for the first time an RTD coupled to a telecom-wavelength operating VCSEL, both systems with proven track record and recognised potential for use in optical neuromorphic systems. In Section \ref{sec:characterisation}, we evaluate the excitable responses of the system, including its well-defined spiking threshold and refractory (lethargy) period. In Section \ref{sec:tasks}, we demonstrate in-device coincidence detection (logical AND) and exclusive logical OR (XOR) tasks, two of the key functionalities for practical operation in networked arrangements. Finally, in Section \ref{sec:model}, we provide future outlook of a nanoscale RTD artificial photonic neuron based on a monolithic design and demonstrate the coincidence detection task in such low footprint, high speed system.
\begin{figure*}[t!]
     \centering
     \includegraphics[width=0.99\textwidth]{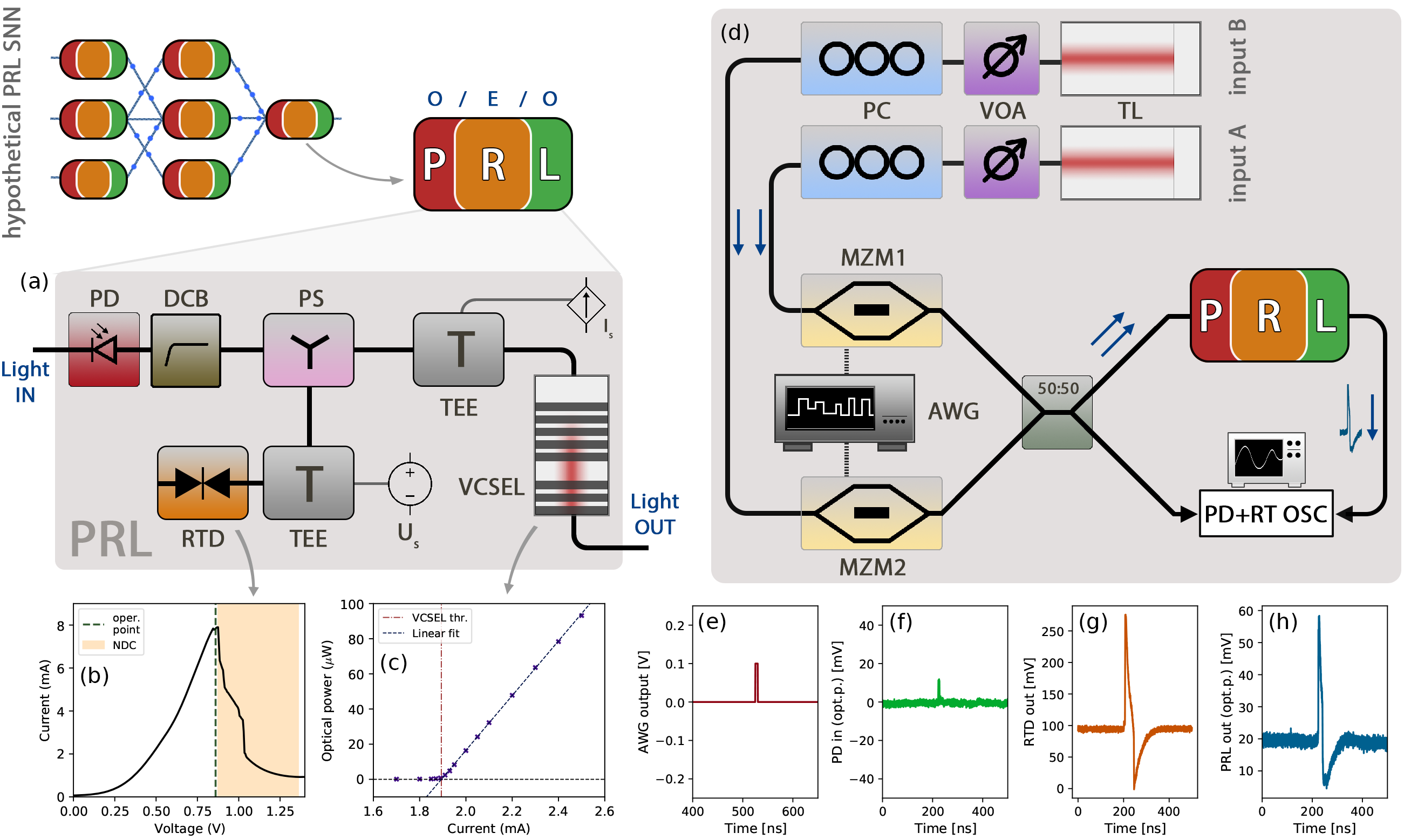}
     \caption{(a) Component layout of the PRL node, including the RTD element with independently tuneable bias $U_s$ coupled to a fiber-coupled VCSEL biased at $I_s$. PD - photodetector; DCB - DC blocking element; PS - resistive power splitter; TEE - a bias tee; RTD - resonant tunnelling diode.
     (b) \textit{I-V} characteristics of the RTD, with region of negative differential conductance (NDC) highlighted in orange. RTD is operated in region around \SI{860}{\milli\volt}. 
     (c) \textit{L-I} curve of the output (transmitter) VCSEL.
     (d) The dual-channel (fan-in) experimental setup with optical timetrace readout on the RT OSC. TL - tuneable laser; ISO - optical isolator; VOA - variable optical attenuator; PC - polarisation controller; MZM - Mach-Zehnder modulator; AWG - arbitrary waveform generator. 
     (e) Example of an spike triggering perturbation (waveform for AWG). 
     (f) Trace of optical power on input branch, RF-modulated via MZM and recorded on amplified photodetector (DC component is filtered). 
     (g) An electrical spike, activated in a PD-RTD system. 
     (h) An optical spike at the output VCSEL.} 
     \label{fig:ExpSetup}
 \end{figure*}
\section{PD-RTD-VCSEL spiking node layout}\label{sec:layout}
The RTD artificial spiking neuron circuit combines an O/E receiver (PD), an excitable RTD and an E/O transmitter laser (a VCSEL). Throughout this work, we will refer to this layout as a \textbf{PRL} node (\textbf{P}hotodetector-\textbf{R}TD-\textbf{L}aser), with the node circuit layout depicted in Fig. \ref{fig:ExpSetup}(a) and experimental layout used to characterise the node shown in Fig. \ref{fig:ExpSetup}(d). The RTD device (with a \SI{3}{\micro\metre} radius circular mesa) was fabricated on a layerstack grown by molecular beam epitaxy on a semi-insulating InP substrate, containing a \SI{1.7}{\nano\metre} AlAs/\SI{5.7}{\nano\metre} InGaAs/\SI{1.7}{\nano\metre} AlAs DBQW structure surrounded by highly doped \textit{n}-InGaAs contact layers. As previously discussed, the RTD exhibits a highly nonlinear I-V characteristic (shown in Fig. \ref{fig:ExpSetup}(b)) with a pronounced NDC region starting around \SI{900}{\milli\volt} and extending beyond \SI{1.25}{\volt} as highlighted with orange shading in Fig. \ref{fig:ExpSetup}(b). Lumped circuit scheme for the RTD optoelectronic circuit can be found in \cite{Ortega-Piwonka2022_OME,Hejda2022_PRAppl}. For operation as an excitable driving element, the RTD is biased via a \SI{12}{\giga\hertz} bias-tee (Inmet 8800SMF1-12) with a DC voltage $U_S$ very close to the peak voltage value (typical operation point highlighted with green dashed line in Fig. \ref{fig:ExpSetup}(b)). When biased near the NDC region, any perturbation of sufficient strength can push the system over the boundary of the NDC, causing the RTD to fire an excitable spike alongside a limit cycle. The \textit{I}-\textit{V} curve of the RTD exhibits a very high peak-to-valley current ratio (PVCR $\approx 8.5$, Fig. \ref{fig:ExpSetup}(b)), which is favourable as it provides spikes with high signal-to-noise amplitude ratio.

The optical input of the PRL node is realised with a \SI{9}{\giga\hertz} amplified InGaAs photodetector (PD, Thorlabs PDA8GS), which directly converts input light perturbations into RF signals that enter the RTD. Unlike in previous works \cite{Romeira2016}, our setup does not require an EDFA to increase input optical power entering the node. The PD is coupled through a DC blocking element (DCB in Fig. \ref{fig:ExpSetup}a) since any offset voltage coming from the PD could shift the RTD operational point from $U_S$. The output of the PRL node is realized with an off-the-shelf VCSEL (Raycan) operating at \SI{1550}{\nano\metre}, driven with an RF-enabled laser mount (Thorlabs LDM56) with thermal control via thermistor ($R$ = \SI{12.5}{\kilo\ohm}) and with an \SI{50}{\ohm} RF input that is AC-coupled directly to the VCSEL through a bias-tee. The lasing threshold of the VCSEL is approx. \SI{1.9}{\milli\ampere} (Fig. \ref{fig:ExpSetup}(c)) at room temperature. The VCSEL output passes through an optical isolator (to avoid unwanted reflections and increase the S/N ratio) and is read out on a 16 GHz real-time oscilloscope (Rohde\&Schwarz RTP) using a second amplified photodetector. The input (PD) and output (RF-IN on the bias-tee driving the VCSEL) terminals ot the PRL node are all connected together via a 2-way-0° resistive \SI{50}{\ohm} power splitter (ZFRSC-183-S+, PS in Fig. \ref{fig:ExpSetup}a). These three main functional blocks constitute the full PRL (O/E/O) photonic spiking neuron. Fig. \ref{fig:ExpSetup}(e-h) shows how a small optical perturbation (Fig. \ref{fig:ExpSetup}(f)) activates a spiking response in the RTD (Fig. \ref{fig:ExpSetup}(g)) and subsequently in the VCSEL output (Fig. \ref{fig:ExpSetup}(h)). In all the Figs., "(opt.p.)" denotes optical power, represented by voltage trace (as produced by the amplified photodetector).

To demonstrate the fan-in functionality and input integration capability for multiple optical upstream signals, we provide two optical inputs (branches) into the PRL node. This showcases that the node can simultaneously process signals at various wavelengths (defined by the sensitivity range of the PD) similarly to other O/E neurons \cite{Tait2017}. Each branch includes an optical isolator (ISO) to limit unwanted reflections, a variable optical attenuator (VOA) for power adjustment and a Mach-Zehnder modulator (MZM) controlled by an arbitrary waveform generator (AWG, Keysight M8190 12 GSa$\cdot s^{-1}$) via a 10dB RF amplifier which provide spike trigger signals. The operational point (bias voltage) of each MZM is set between the quadrature and the maximum of the output power transfer curve to achieve near-linear relation between input amplitudes and output light intensity modulations. The two independent branches are then combined via a 50:50 fibre-optic coupler and fed into the input (PD) of the PRL node. We denote the input branches as: \textbf{A} and \textbf{B}. Branch A utilizes a tunable laser (Santec TSL-210) operating at \SI{1310}{\nano\metre}. The MZM (Thorlabs) in branch A is biased at \SI{7}{\volt}. Branch B utilizes another tunable laser (Santec WSL-110) operating at \SI{1546}{\nano\metre} and the MZM (JDS Uniphase) is biased at \SI{2.1}{\volt}. The average CW power provided from each branch was approx. \SI{350}{\micro\watt}.

We want to emphasise that this off-the-shelf realisation of the PRL neuron is a first proof-of-concept demonstration of an RTD-based O/E/O spiking neuron with deterministic spike triggering. In the future, an optically-sensitive RTD (with an embedded optical input window) would enable for co-integration of the PD-RTD elements into single (sub)-micron sized structure. In parallel, a fully monolithic layout incorporating all the active components into a single structure is under investigation \cite{Romeira2021}. These monolithic designs \cite{Hejda2022_PRAppl, Ortega-Piwonka2022_OME} are of significant interest, as they would significantly reduce the footprint of the PRL nodes down to sub-micrometric dimensions and allow for their chip-scale integration. Furthermore, we also want to highlight that the spiking rates of the proposed O/E/O PRL node are mostly governed by the RTD's circuit parameters, and not limited by the RTD structure itself since the resonant tunnelling effects in the RTDs persist beyond GHz rates. Therefore, future optimisation of circuit parameters, in addition to transition towards monolithically-integrated designs, will enable optoelectronic spiking systems yielding ultrafast spiking (multiple GHz) rates. The performance of such future monolithic integrated PRL node is investigated numerically at the end of this work in Section \ref{sec:model}.

\section{PD-RTD-VCSEL spiking node characterisation}\label{sec:characterisation}

For a system to operate as a spiking artificial neuron, multiple functional requirements need to be fulfilled. These include the ability to receive inputs (ideally from multiple upstream nodes), perform input signal summation and thresholding, and finally respond with all-or-nothing responses that are generally independent on the shape of the activating perturbation. The excitable spiking neuron should also act as a filter for perturbations arriving too soon after the firing of a spike event (thus exhibiting refractoriness), and should also integrate (summate) multiple sub-threshold pulses arriving at the same time. We demonstrate experimentally that the PRL node fulfills all these requirements. 

First, a spiking artificial neuron should exhibit a clear excitability threshold, upon which an all-or-nothing response (spike) is fired. A single sub-threshold input perturbation (stimulus) should not result in any significant response from the system. Conversely, a super-threshold perturbation should cause the firing of a spiking event, where the shape of the spike is fully governed by the dynamical limit cycle (i.e. the circuit, not by input perturbation's shape).

 \begin{figure}[ht]
     \centering
     \includegraphics[width=0.45\textwidth]{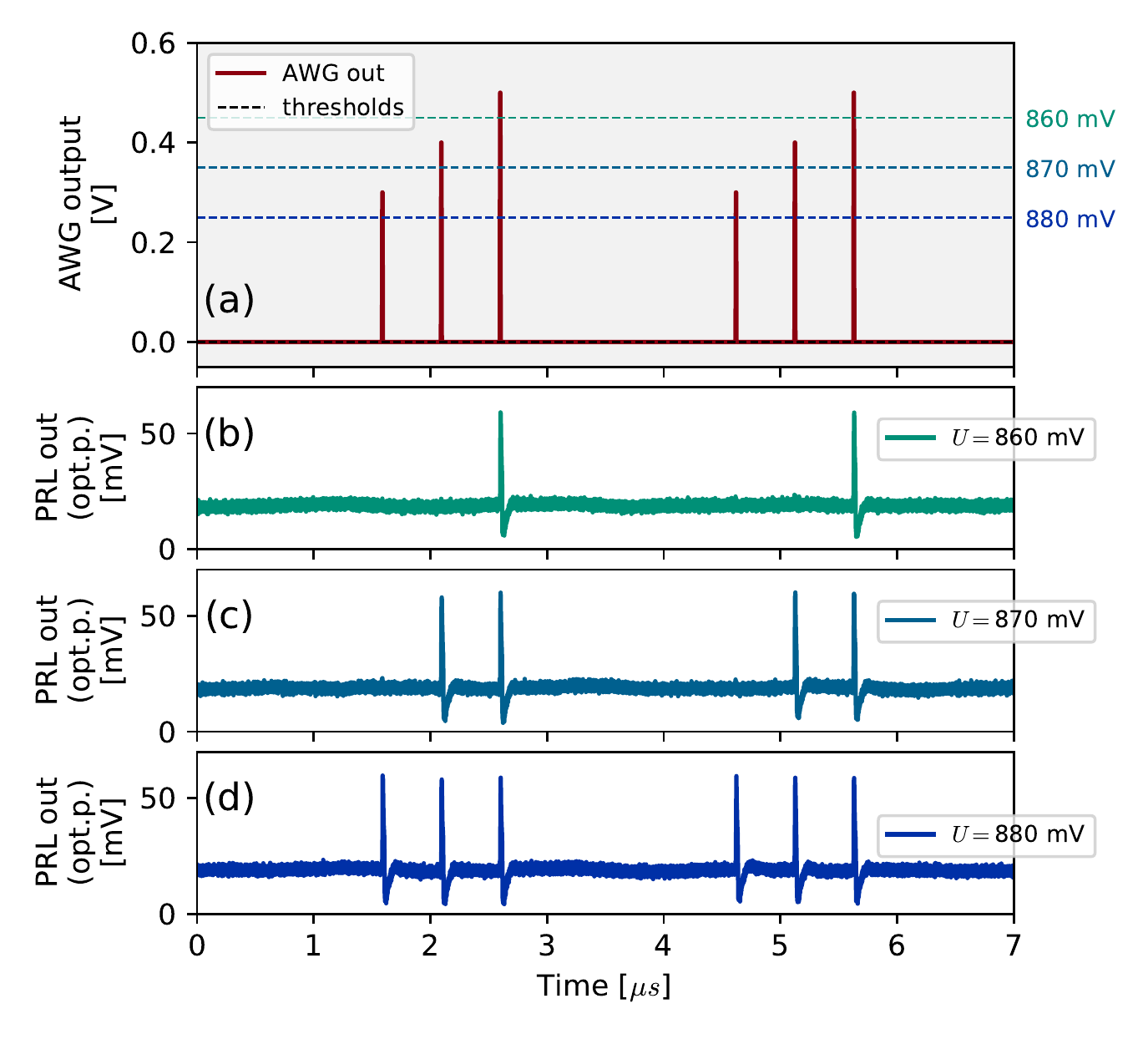}
     \includegraphics[width=0.38\textwidth]{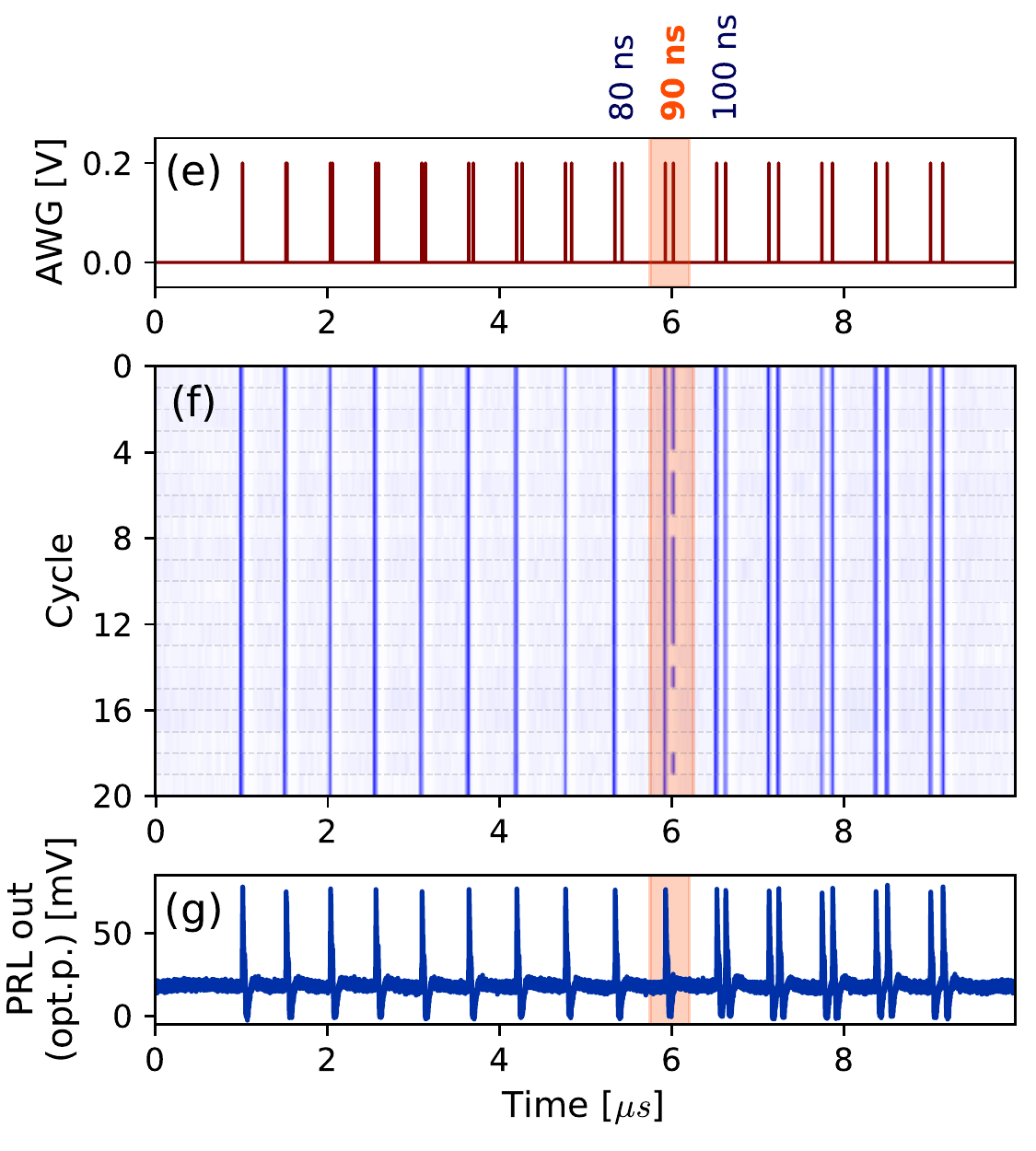}
     \caption{Threshold and refractoriness characterization in the PRL. 
     (a) Source waveform of stimuli (\SI{3}{\nano\second} square pulses) with varying amplitude. This serves as PRL input via MZM on the input branch to the PRL. Dashed lines show approximate threshold levels for different RTD bias voltages. (b)-(d) Timetraces of VCSEL output as recorded on an amplified photodetector. For the lowest bias voltage value (b) $U=$ \SI{860}{\milli\volt}, the threshold is further from the steady state and only the strongest perturbation triggers a response (a spike). Response for same signal with RTD bias $U=$ \SI{870}{\milli\volt} (c) and response with RTD biased closest to the threshold, for $U=$ \SI{870}{\milli\volt} (d); Refractory period demonstration with temporal map ($n=20$ cycles).; Refractory period study: (e) The source modulation, consisting of pulse pairs with gradually increasing temporal separation (\SI{10}{\nano\second} increments).
     (f) The temporal map shows the traces of 20 subsequently recorded measurement cycles of \SI{10}{\micro\second} length. (g) Example of one response trace of the PRL node.} 
     \label{fig:thresholding_and_tempmap}
 \end{figure} 
     
Fig. \ref{fig:thresholding_and_tempmap}(a-d) demonstrates the clear thresholding behaviour in the PRL node, showing how variations of the RTD bias voltage $U$ shift the distance of the threshold from the quiescent state. In this experiment, a sequence of three ($t_{sep}=$ \SI{500}{\nano\second}) square-shaped unipolar \SI{3}{\nano\second} pulses of varying amplitudes is used to modulate the optical signal from one of the input branches. The system responds in an all-or-nothing manner, with sub-threshold pulses not causing any observable response at the PRL output trace. It can also be seen that increasing the $U$ value from \SI{860}{\milli\volt} to \SI{880}{\milli\volt} moves the excitable threshold closer to the steady-state, resulting in spike-activation for input pulses with lower amplitudes. 

Another hallmark characteristic of excitable systems is the presence of a refractory (lethargic) time. After a spiking response is elicited within the system, it exhibits a time period in which it is unable to respond with another full response. In certain classes of artificial neuron models, a distinction can be made between an absolute refractory time (period during which the system can not respond at all) and a relative refractory period (where the probability of spike activation is reduced or amplitude of the spiking response is diminished). 
Fig. \ref{fig:thresholding_and_tempmap}(f,g) demonstrates that the PRL node exhibits a clearly defined refractory period. To test this, we injected input perturbations into the PRL in the form of a sequence of super-threshold \SI{3}{\nano\second} pulse doublets with gradually increasing temporal separation between the individual pulses in each pair. Temporal separation values between $\Delta_t=$ \SI{0}{\nano\second} and $\Delta_t=$ \SI{150}{\nano\second} have been used with \SI{10}{\nano\second} increments. To better evaluate the refractory period of the system, 20 subsequent oscilloscope readouts (cycles) have been acquired and processed into a temporal map shown in Fig. \ref{fig:thresholding_and_tempmap}(f). The orange highlight shows the pulse doublet separation for which a second excitable event starts occurring, directly demonstrating the refractory period of the system $T_{ref} \approx $ \SI{90}{\nano\second}. Due to the presence of noise and temporal jitter, the observed spike activation probability per separation is not a strict step-like function, but exhibits an interval (here $\Delta_t=$ \SI{90}{\nano\second}) where the likelihood of a follow-up spike is reduced. Since the upper limit of excitable system's spiking rate is governed by the refractory time, we can define the maximum spiking frequency for the PRL as $\frac{1}{T_{ref}}\approx$ \SI{10}{\mega\hertz}. As the temporal map in Fig. \ref{fig:thresholding_and_tempmap}(f) also demonstrates, the spike excitation process via super-threshold pulses offers very high degree of reliability, with no missing spikes observed in the 20 recorded cycles.
 
\section{PD-RTD-VCSEL spiking node functional tasks}\label{sec:tasks}

Artificial neurons in neural networks rarely process information from a singular upstream source. Therefore, having the many-to-one fan-in functionality (processing multiple simultaneous inputs) is key for realisation of larger-scale neural networks. It is generally assumed that signal coincidences and synchrony play a major role in the brain \cite{Agmon-Snir1998} and in sensory pathways, including the auditory \cite{Stevens1998} cortex. Among other functionalities, coincidence detection enables for mirror symmetry density detection \cite{George2021_SciRep}, which is vital for perception procedures based on visual stimuli in image processing algorithms. Furthermore, for propagation of information in larger neural networks, it is also important to have the capability of information subtraction. An example of this is the exclusive OR (XOR) logic task, where each individual input coming from one of the two branches results in a \textit{TRUE} state, while both pulses coming together result in a \textit{FALSE} state in the system. In this section, we experimentally demonstrate these two fundamental tasks, performed with a pair optical inputs entering into a single PRL node: a coincidence detection (logical \textit{AND}) and an exclusive or (logical \textit{XOR}) task.

\begin{figure}[ht]
     \centering
     \includegraphics[width=0.455\textwidth]{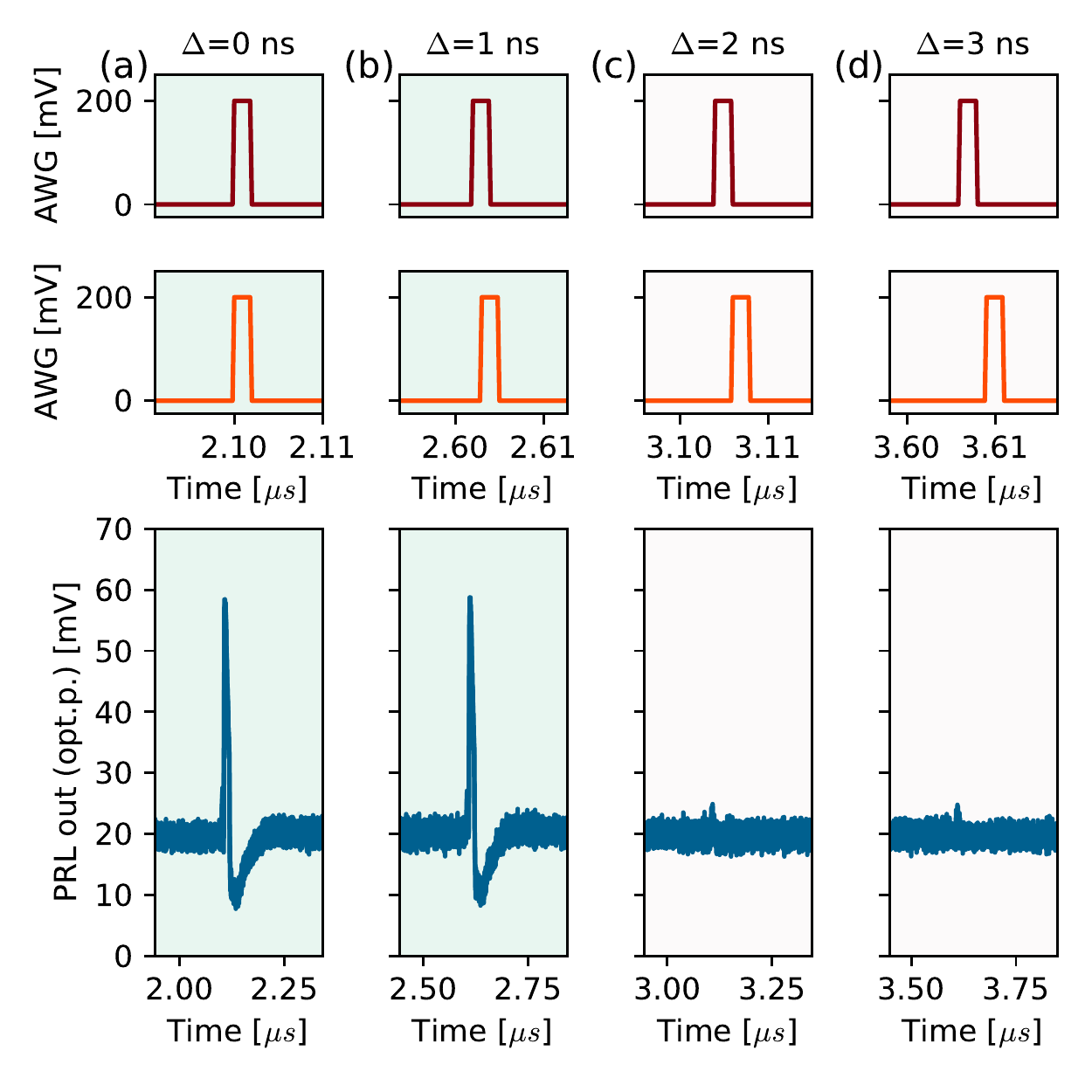}
     \includegraphics[width=0.535\textwidth]{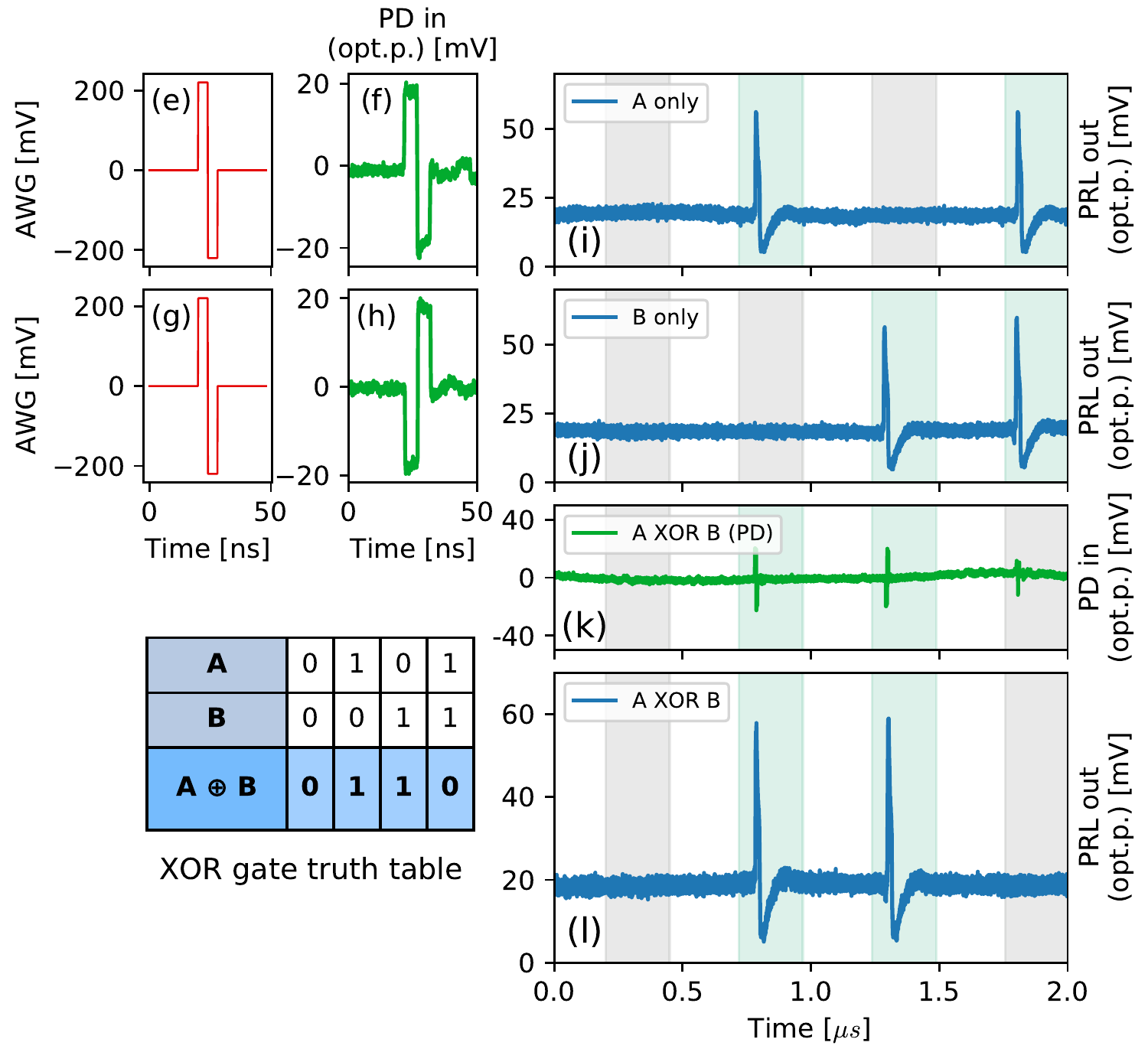}
     
     \caption{Two-input information processing tasks in the PRL. (a-d) Coincidence detection (logical AND), where the first two rows show the \SI{2}{\nano\second} square pulses (stimuli) gradually shifting apart in time. For (a) perfect and for (b) partial overlap, the total energy at a given moment surpasses the firing threshold, eliciting a spike. However, as the pulses drift further apart (c,d), the input perturbation energy to the system at any given time is not sufficient for spiking.; (e-l) Demonstration of an exclusive OR (XOR) in the PRL: (e,g) Bipolar RF-modulation trigger pulses, (f,h) MZM-output for each branch, together providing input to the PRL. The input pulses can individually elicit spiking responses from the PRL node (i,j), but when arriving simultaneously, they (k) cancel out and no spike is fired in such instance (l). This demonstrates the XOR task performance. The truth table for the task is included for reference.}
     \label{fig:coinc_and_xor}
\end{figure}


Fig. \ref{fig:coinc_and_xor} shows the results for both of the tasks. First, we demonstrate pulse coincidence detection performed on two independent optical inputs entering the PRL node (Fig. \ref{fig:coinc_and_xor}(a-d)). Both input branches were modulated with \SI{2}{\nano\second} square pulses with gradually increasing mutual temporal separations, with increments (measured between rising edges of pulses) of $\Delta_{incr}$ = \SI{1}{\nano\second}. As can be seen in Fig. \ref{fig:coinc_and_xor}(a,b), only perfectly overlapping pulses $\Delta_{incr}$ = \SI{0}{\nano\second} and partially overlapping pulses $\Delta_{incr}$ = \SI{1}{\nano\second} result on total sum of optical input power exceeding the excitability threshold, leading to the triggering of a single excitable spike in the RTD. For separations $\Delta \geq$ \SI{2}{\nano\second}, no responses are triggered. This functionality can be considered as a temporal version of the \textit{AND} logical gate. 
Next, we demonstrate the XOR task (Fig. \ref{fig:coinc_and_xor}(e-l)), with XOR truth table added for reference. 
In this case, the two input branches use equivalent, bipolar, spike-like modulation signals (red trace, Fig. \ref{fig:coinc_and_xor}(e,g)). These perturbations are converted to optical signals using MZMs (Fig. \ref{fig:coinc_and_xor}(f,h)). Here, each MZM is set in opposing points of the transfer function between quadrature and maximum, resulting in polarity inversion for the pulses in branch B (Fig. \ref{fig:coinc_and_xor}(h)). These bipolar optical pulses are then propagated to the PD element in the PRL node. Each pulse on its own is capable of eliciting an individual spike, since the positive part of the perturbation exceeds the spike activation threshold of the RTD (Fig. \ref{fig:coinc_and_xor}(i),(j)). However, when both of these pulses arrive on the PD element of the PRL node simultaneously, they cancel each other out (Fig. \ref{fig:coinc_and_xor}(k)) as they are mutually in anti-phase. As a result, no spike is elicited in the PRL node for active A and B (Fig. \ref{fig:coinc_and_xor}(l)). Therefore, the full set of required states for the XOR task is achieved by the combination of optical signal summation, thresholding and optical spike firing in the PRL system.

\section{Perspective: towards nanoscale RTD-based optoelectronic spiking nodes\label{sec:model}}
Following the experimental results on the system's refractoriness (Fig. \ref{fig:thresholding_and_tempmap}), the lower bound on the time-interval between two consecutively fired spiking events is approx. \SI{100}{\nano\second}. This already allows for spiking rates at up to \SI{10}{\mega\hertz} with this proof-of-concept experimental system demonstration. With further optimizations, we believe higher speeds are attainable since RTD oscillator circuits can operate at much faster frequencies, typically governed by the circuit's layout and parameters, e.g. (parasitic) inductance of the circuit $L$ and capacitance $C$, among other factors. Simultaneously, it is of vital importance to explore the prospects of component integration and spatial footprint reduction.

\begin{figure*}
     \centering
     \includegraphics[width=0.9\textwidth]{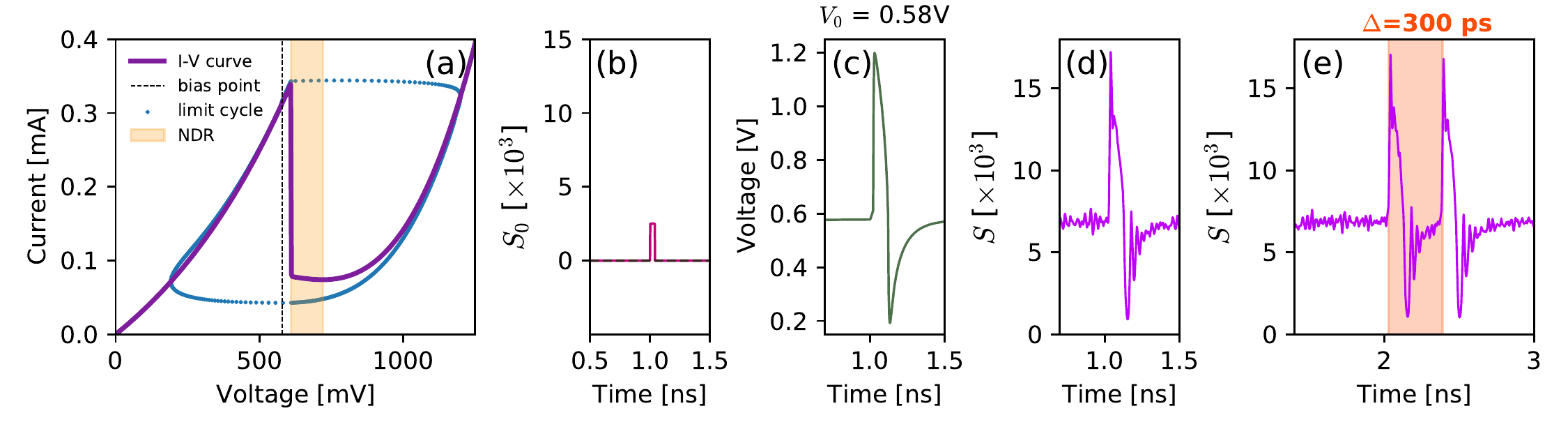}
     \caption{Numerical simulation of a PD-RTD-LD system. (a) I-V characteristic of a nano-scale RTD, with highlighted NDC region, voltage bias (fine dashed line) and evolution of the system states during the spiking event, showing the dynamical limit cycle. (b) Example of an electrical trigger signal. (c) Electrical response of the nano-RTD. (d) Optical response of a voltage-coupled nanoscale RTD-LD system. e) Demonstration of firing of a pair of consecutive optical spikes, elicited in the system when subject to the arrival of a pair of input perturbations with temporal separation of \SI{300}{\pico\second}.}
     \label{fig:sim}
\end{figure*}

To highlight the feasibility of RTD optoelectronic circuits for high-speed spiking functionalities with highly-reduced footprints, we investigate numerically the operation of a nanoscale monolithic PRL node comprising in its structure an RTD capable of operating at beyond GHz spiking rates. Similarly to the experimental results on the PRL system built with the off-the-shelf components, the numerical model investigating the operation of a nanoscale-PRL node combines three main functional elements: a voltage-coupled photodetector term; a two-dimensional system of ordinary differential equations describing the evolution of $I$ and $V$ in a nanoscale RTD-based circuit; and a two-dimensional nanolaser model with an included stochastic term. The dynamics of the nanoscale PRL are modelled using a set of ordinary differential equations:

\begingroup
\allowdisplaybreaks
\begin{align}
C\frac{dV}{dt} & =I-f(V)\label{eq:RTD_V}\\
L\frac{dI}{dt} & =V_0 + R\kappa S_{0}(t)-V-RI\label{eq:RTD_I}\\
\frac{dS}{dt} & =\left(\gamma_{m}(N-N_{0})-\frac{1}{\tau_{p}}\right)S+\gamma_{m}N+\sqrt{\gamma_{m}NS}\xi(t)\label{eq:LD_S}\\
\frac{dN}{dt} & =\frac{\eta V}{q_{e}R_0} -(\gamma_{l}+\gamma_{m}+\gamma_{nr})N-\gamma_{m}(N-N_{0})|E|^{2}\label{eq:LD_N}
\end{align}
\endgroup
where $V$ is the voltage along the RTD, $I(t)$ is the total current in the circuit, $S(t)$ is the photon number and $N(t)$ is the carrier number. 
In difference to previous works \cite{Romeira2013}, the RTD (first two Eqns.) and laser (last two Eqns.) models are coupled through the voltage term to represent same behaviour as observed experimentally, where RF signals are AC-coupled directly to the laser through a bias-tee network (here, $R_0$ = \SI{50}{\ohm} and coupling efficiency $\eta$ = 0.2). The PD element is coupled as a voltage modulation via the $R \kappa S_{0}(t)$ term, where $\kappa$ represents the opto-electrical conversion factor of the PD (sensitivity, in units of current per photon count), hence defining $R \kappa$ as the detector conversion gain.
The $I$-$V$ characteristic (Fig. \ref{fig:sim}(a)) of the RTD is based on a physical model by Schulman \cite{Schulman1996}. The curve parameters, based on experimental measurement of nano-scale RTD devices, as well as the laser and circuit parameters are available from \cite{Hejda2022_PRAppl}. Using these $I$-$V$ parameters yields a highly nonlinear $I$-$V$ curve with an NDC region between \SI{609}{\milli\volt} and \SI{720}{\milli\volt}.
As shown in Fig. \ref{fig:sim}, an incoming optical perturbation in the form of a \SI{20}{\pico\second} long pulse (Fig. \ref{fig:sim}(b)) is sufficient to trigger an ultrafast (sub-ns) spiking response with amplitude of over \SI{1}{\volt} peak-peak (Fig. \ref{fig:sim}(c)). Direct comparison with the traces displayed in Fig. \ref{fig:ExpSetup}(e) confirms very good agreement between the dynamics observed experimentally with the macroscopic PRL node (PD-RTD-VCSEL design) and numerically for the modelled nanoscale PRL system. This signal is then fed into the nanolaser within the nano-PRL node's structure, resulting in high speed optical spikes (Fig. \ref{fig:sim}(d)). In addition to the consistency between the numerical and experimental dynamics, we must note that the nanoscale PRL system allows for significantly improved spiking rates, currently projected as approximately three magnitudes faster than those obtained in our experiments with the PD-RTD-VCSEL based PRL system. To obtain the estimate of upper limit on the spiking rate of this nanoscale PRL model, we performed a spiking refractoriness evaluation by using pairs of pulses with gradually increasing temporal separation (the same test that was carried out experimentally in Fig. \ref{fig:thresholding_and_tempmap}). The achieved lowest temporal separation between two input pulses that reliably elicited a pair of spiking responses from the nanoscale PRL node was \SI{300}{\pico\second} (Fig. \ref{fig:sim}), resulting in a maximum theoretical spiking rate of \SI{3.3}{\giga\hertz}.
\begin{figure}[ht]
     \centering
     \includegraphics[width=0.5\textwidth]{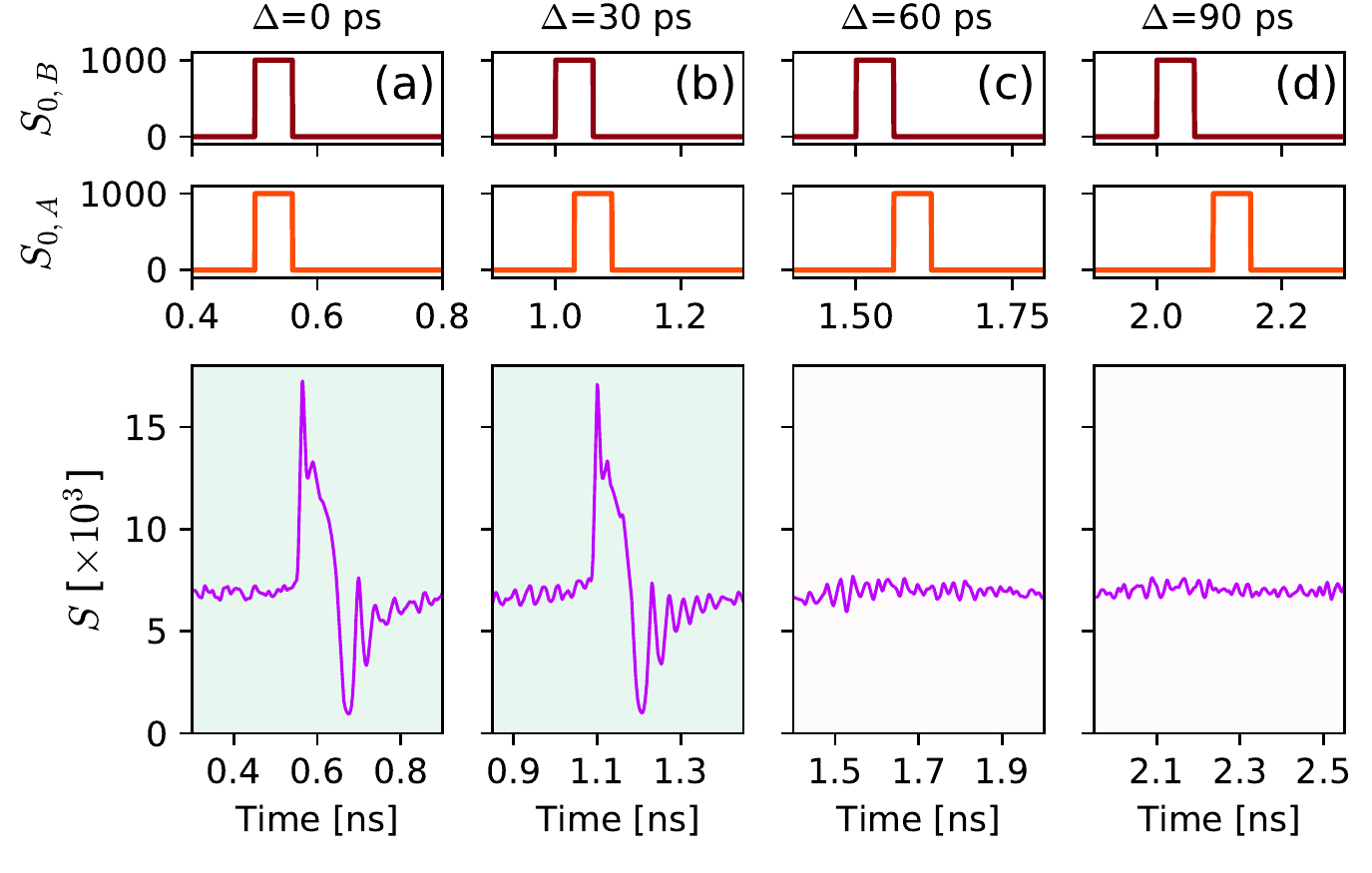}
     \caption{Numerical validation of the two-pulse coincidence detection task. Two positive square pulses encoded in the optical input signal trigger an excitable response in the PRL model only when both arrive into the node at the same ($\Delta =$ \SI{0}{\pico\second}, (a)) or nearly the same time ($\Delta =$ \SI{30}{\pico\second}, (b)). Pulse pairs with larger temporal separation (c,d) do not trigger an excitable response in the modelled nanoscale PRL node.}.
     \label{fig:sim_coinc}
\end{figure}
Furthermore, we validate in theory that the same coincidence detection functionality (as demonstrated experimentally in Fig. \ref{fig:coinc_and_xor}) can be achieved with the nanoscale PRL node, yet again at significantly faster rates. In Fig. \ref{fig:sim_coinc}, a pair of \SI{60}{\pico\second} optical pulses of amplitude $S_0 = $ 1000 is incident on the PD element of the nano-PRL node. For either no temporal shift ($\Delta = $ \SI{0}{\pico\second}, Fig. \ref{fig:sim_coinc}(a)) or small temporal shifts preserving some overlap between the input pulses (($\Delta = $ \SI{30}{\pico\second}, Fig. \ref{fig:sim_coinc}(b)), the system performs coincidence detection by firing a sub-ns spike. Higher separation times between input pulses do not elicit a spiking response (Fig. \ref{fig:sim_coinc}(c,d)). Therefore, the model confirms the functionality observed in the experimental PRL node also for the monolithically-integrated nanoscale PRL node and at higher processing speeds. 
\section{Conclusion}
This work demonstrates experimentally a RTD-based optoelectronic circuit operating as an excitable, spiking artificial neuron, benefiting from optical I/O ports, allowing its use in prospective optically interlinked spiking photonic neural networks. The system, referred as PRL node, combines in its structure a micrometric InGaAs/AlAs RTD element with high PVCR $\approx 8.5$ coupled to a photodetector and a telecom-wavelength VCSEL, providing respectively the optical input and output of the PRL node. We have demonstrated how such system allows for multiple optical pulsating (spiking) input signals (here $n=2$) to be summed up on the photodetector of the node in a wavelength-independent manner, allowing for additional robustness in operation and lower demands on coherence and precise wavelength selection in future interconnected systems based upon interlinked PRL nodes. We have also shown the presence of a well-defined excitability threshold in the system, and a spiking refractory period of $T_{ref}\approx$ \SI{90}{\nano\second} in the proof-of-concept system realisation of this work. This allows already, without any additional system optimisation stage, a reliable activation of all-or-nothing spiking responses at up to \SI{10}{\mega\hertz} rates. By further optimization of RTD's electronic circuit and by progressing towards monolithically-integrated structures with highly reduced dimensions, we believe the operation speed can be significantly increased. We validate this claim numerically, showing that the experimentally achieved spiking functionalities can be also obtained with a nanoscale RTD-based O/E/O system at GHz rates. Finally, we have successfully demonstrated spike-based processing tasks with the proposed PRL node, including a coincidence detection (logical AND) and an exclusive OR (XOR) task. These results provide the first investigation and proof-of-concept demonstration of an O/E/O RTD-based, deterministically activated spiking photonic neuron. Future research will focus on system optimisation towards faster spiking rates reaching GHz speeds, as well as on the development of improved RTD devices with embedded optical windows to allow direct photodetection in the RTD structure; thus permitting to eliminate the currently used externally-coupled PD module. Additionally, we will focus on experimental validation of signal propagation between master-receiver PRL nodes, ultimately aiming towards photonic spiking neural networks powered by these artificial excitable optoelectronic neurons.

\begin{acknowledgement}

The authors would also like to acknowledge J. Iwan Davies from IQE for providing the epitaxial growth for RTD fabrication.
\end{acknowledgement}

\begin{funding}
The authors acknowledge support by the European Commission (Grant 828841-ChipAI-H2020-FETOPEN-2018-2020) and by the UK Research and Innovation (UKRI) Turing AI Acceleration Fellowships Programme (EP/V025198/1). 
\end{funding}
\printbibliography
\end{document}